\documentclass[aps,prl,twocolumn,groupedaddress]{revtex4}
\usepackage{graphicx}

\begin{document}


\title{Pion Interferometry from pp and dAu Collisions at RHIC}


\author{Thomas D. Gutierrez}

\affiliation{University of California, Davis \\
For the STAR Collaboration}


\begin{abstract}
Pion interferometry results from pp and dAu collisions at $\sqrt{s}=200$ GeV are
presented using data from the STAR detector at RHIC.
Radii from three-dimensional
correlation functions are studied as a function of transverse mass.
The HBT results from pp and dAu collisions are compared to pion HBT from AuAu reactions, 
also taken at STAR.
\end{abstract}


\maketitle


Hanbury-Brown Twiss interferometry has long been used as a tool to characterize the
space-time freeze-out geometry of particles produced in elementary particle and heavy ion collisions \cite{hbtgold60,hbtgyu79,hbtboal90,hbthei99}.  At
RHIC (Relativistic Heavy Ion Collider), there is a unique opportunity to study pion HBT in 
three separate systems, pp, dAu, and AuAu, using the same
detector at the same center of mass energy.  

In this work, presented as a poster at Quark Matter 2004, pion HBT results from pp and 
dAu collisions are compared to AuAu results 
all taken at $\sqrt{s}=200$ GeV using the STAR (Solenoidal Tracker at RHIC) detector.  
Of primary interest in this report is to compare how the three-dimensional
regions of homogeneity vary with the transverse mass of the pion pairs in each system
($M_T=\sqrt{m^2+k_{T}^2}$ where $k_T$ is the pair's transverse momentum).  

The central observable in HBT is the two particle correlation function, $C_2$.  This correlation function is sensitive
to several interesting physics effects: the quantum statistics obeyed by the pairs being studied (Boson, Fermion, etc.), 
the quantum field configuration (thermal,
coherent, etc.), the source geometry, the source dynamics (space momentum correlations, flow, jets, etc.), and pairwise interactions 
(Coulomb, strong, etc.).
The broader goal of HBT studies in hadronic physics 
is to isolate and understand these individual effects in various systems under different conditions.

Experimentally, the two-pion correlation function is obtained from the ratio
$C_2(\vec{q})=A(\vec{q})/B(\vec{q})$.  The quantity $A(\vec{q})$ is the two-pion distribution 
of pair momentum differences, $\vec{q}=\vec{p}_1-\vec{p}_2$.  The mixed background,
$B(\vec{q})$, is the two-pion distribution generated from pion pairs coming from different events.

The correlation function was measured in three dimensions using the Bertsch-Pratt decomposition of $\vec{q}$:
$q_{\rm{out}}$, $q_{\rm{side}}$, and $q_{\rm{long}}$.  Here, $q_{\rm{long}}$ is parallel to the beam.
The component $q_{\rm{side}}$ is
perpendicular to both $q_{\rm{long}}$ and to the transverse momentum vector of the pair, $\vec{k}_T$.  The third
component, $q_{\rm{out}}$, is perpendicular to both
$q_{\rm{long}}$ and $q_{\rm{side}}$.  The measured correlation function was then fit to a Gaussian

\begin{equation}
C_2(q)=1+\lambda e^{-R_{\rm{out}}^{2} q_{\rm{out}}^{2}-R_{\rm{side}}^{2} q_{\rm{side}}^{2}-R_{\rm{long}}^{2} q_{\rm{long}}^{2}}
\label{c3d}
\end{equation}
where $\lambda$ represents the strength of the correlation at zero momentum difference.  It is also commonly known as the 
``chaoticity parameter.''  The correlation function was also studied in one-dimension
versus 
$Q_{\rm{inv}}=\sqrt{|(E_1-E_2)^2-(\vec{p}_1-\vec{p}_2)^2|}$.  Here, the correlation function was parameterized with the Gaussian
\begin{equation}
C_2(Q_{\rm{inv}})=1+\lambda e^{-R^{2} Q_{\rm{inv}}^2}.
\label{cqinv}
\end{equation}

The primary source of systematic error in the pp and dAu systems comes
from the assumption about the shape of the correlation function.  Future work will include extensively characterizing the correlation
function using other methods such as the Edgeworth expansion, Legendre expansion \cite{hbtcso00}, and imaging \cite{hbtbro97}.  Systematic error on the HBT parameters
are estimated to be as large as 20\% using a Gaussian shape assumption.  In spite of this, 
the Gaussian shape still 
reasonably characterizes the global properties of the correlation function for purposes of comparison between the systems.  

The primary subdetector used at STAR in this analysis is the main Time Projection Chamber (TPC), described 
elsewhere \cite{hbtstar01}.  Particle identification (PID) was achieved by correlating the magnetic rigidity of a track with its 
specific ionization ($dE/dx$).  For the pp and dAu analyses, a one sigma cut was made around the pion Bethe-Bloch band while
the electrons, kaons, and protons were suppressed at the two-sigma level providing a pure sample of charged pions.  The contamination 
due to electrons and kaons is small and contributes to a slight systematic error of about 1\%, an estimate based on changing
the width of the sigma cuts.  Good pion PID was achieved to a momentum of about 800 MeV and the TPC itself imposes an lower momentum
cut at 100 MeV.  Pion tracks were selected at mid-rapidity, $|y|<0.5$, for this analysis.

Before cuts, approximately 10 million minimum bias pp events and 8 million minimum
bias dAu events were used.  

Two sources of pairwise systematic error, track splitting and track merging, were addressed.  Track splitting occurs when
a single track is reconstructed as two in the TPC and track merging is when two tracks are mistaken as one.  Both
effects can affect the shape of the correlation function at very low $q$.  In AuAu collisions, the correlation function is quite
narrow because the pion source size is large ($\sim 6$ fm).  
For that system, this makes eliminating splitting and merging particularly important.  
However, in dAu and pp collisions, the source
is comparatively small, giving a wide correlation function.  In pp collisions at STAR, splitting and merging effects 
only significantly affected the first 30 MeV/c of a wide correlation function ($\sim 200$ MeV/c).  However, for future studies where
a detailed understanding of low-q behavior will be explored, removal of these systematic errors 
will become increasingly important and so were implemented in this study for both the pp and dAu systems for completeness.  

A standard pairwise Coulomb correction was applied, similar to that done previously at STAR \cite{hbtstar01}.  This standard
Coulomb correction has been shown to over-correct the correlation function in heavy ion physics
and is inadequate in that context.  However, the overall effect of this standard Coulomb correction 
is small for small sources.  It typically affects only the first 40 MeV/c of the correlation function.  
The use of other correction schemes does not significantly affect the trends of 
the extracted Gaussian parameters for pp and dAu collisions.  
However, for future detailed shape studies, the Coulomb correction will become extremely important. Ideally, one would like to move away from
implementing Coulomb corrections altogether and employ methods such as imaging \cite{hbtbro97} that incorporate all appropriate final state interactions
into the extraction of the source function itself.

The Bowler-Sinyukov-CERES scheme is currently the Coulomb correction method of choice for heavy ion HBT at STAR \cite{hbtlisa03}.  However,
the appropriateness of this scheme in pp and dAu collisions is still being studied.  For example, the Bowler scheme assumes that 
the source is perfectly chaotic and any deviation of the chaoticity parameter, $\lambda$, from unity 
is the result of contamination, long lived resonances, or any other undesirable effects
that artificially reduce the correlation strength.  However, this assumption may not be appropriate for pp collisions where
coherence and core-halo considerations could be important.

\begin{figure}[]
\resizebox{.4\textwidth}{!}{\includegraphics{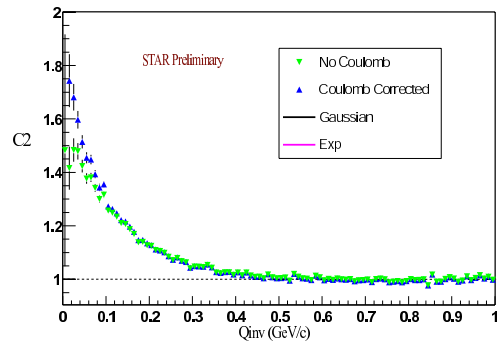}}
\caption{\label{onedcor} A typical one-dimensional $\pi\pi$ correlation function from a pp collision
at $\sqrt{s}=200$ GeV/$c^2$ at STAR.  Both Gaussian and exponential fits to the Coulomb corrected data are shown.  
The darker points are the Coulomb corrected data.
}
\end{figure}

Figure~(\ref{onedcor}) shows a one-dimensional $\pi\pi$ correlation function for a pp collision at STAR with $dN/d\eta=4$, representing
a typical event.  The darker data points, as indicated on the figure, are Coulomb corrected.  Both a Gaussian and an exponential were fit the to the
Coulomb corrected data.  The corrected correlation function is clearly not Gaussian.  As a guideline, 
the width of the correlation function in relative
momentum space is inversely related to source size.  For an incoherent source of identical, spinless bosons, $C_2(q=0)$ should approach 2.  This in turn
should be reflected in the parameter $\lambda$, which should approach unity for a fit functions of the form 
of Eq.~(\ref{cqinv}).  Because of
experimental effects (such as PID contamination) and physics effects (such as coherence and resonances) the value of $C_2(q=0)$ is
often measured to be less than 2.

\begin{figure}[]
\resizebox{.4\textwidth}{!}{\includegraphics{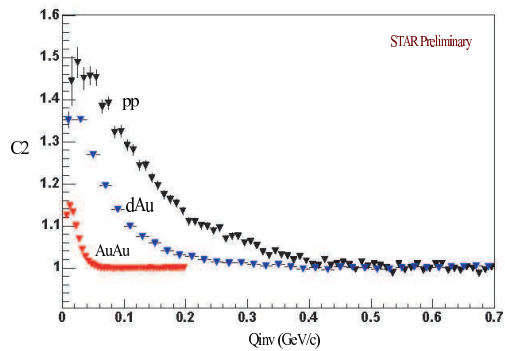}}
\caption{\label{onedall} Typical one-dimensional $\pi\pi$ correlation functions for the pp (upper), dAu (middle), 
and AuAu (lower) systems at STAR.  The pp and dAu correlation functions are from minimum bias data while the 
correlation function from AuAu is shown for central collisions. All are subject to the same $k_T$ cut.}
\end{figure}

It is interesting to compare $C_2(q)$ between different systems.  Figure~(\ref{onedall}) shows three $\pi\pi$ 
correlation functions versus $Q_{\rm{inv}}$ for pp (minimum bias), dAu (minimum bias), and AuAu (central) 
that have not been Coulomb corrected. 
As the system size decreases from AuAu to pp, the width of the correlation function increases.  Note that Bose-Einstein correlations are not
a small effect in dAu or pp collisions in comparison to AuAu.  

\begin{figure}[]
\resizebox{.4\textwidth}{!}{\includegraphics{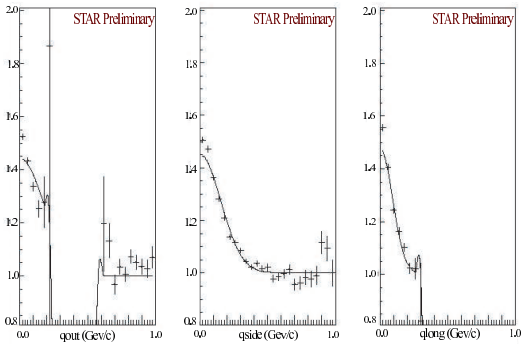}}
\caption{\label{threedpp} Typical out, side, and long projections of the
three-dimensional $\pi\pi$ correlation functions for pp collisions at STAR.
The one-dimensional projections shown are 80 MeV/c in the other directions.  The origin of the
hole seen in the out direction is discussed in the text.}
\end{figure}

Figure~(\ref{threedpp}) shows one-dimensional projections of a three-dimensional 
Bertsch-Pratt correlation function along the out, side, and long directions for pp collisions.  The projections are
80 MeV/c wide in the ``other'' directions for each plot.  The hole, as seen in the out direction,
is a kinematic effect.  It is the result of a pairwise vector $k_T$ cut ($0.15<k_T<0.25$ MeV/c) in combination with an implicit single particle $p_T$ 
cut ($p_T>0.1$ MeV) due to acceptance in the TPC.  

The $M_T$ dependence of the three dimensional HBT parameters provides information 
about the space-momentum correlations present in each system.  A transverse momentum dependence of the longitudinal radius along the beam
axis is associated with a boost-invariant expansion.  Such behavior is commonly seen in a variety of
high energy physics experiments where HBT parameters are studied over a boost-invariant region of longitudinal phase
space.  Other space-momentum correlations, such as longitudinal flow in heavy ion collisions, 
become augmentations to this intrinsic dependence.  A transverse momentum dependence 
of the transverse HBT radii is indicative of an explosive expansion of the pion source, usually associated with
global collective properties such as transverse flow.  In elementary particle collisions, a transverse momentum dependence of the transverse
radii have alternatively been associated with jets, string tension, and Heisenberg uncertainty quantum limits but not 
with collective final-state space-momentum correlations. 
The dAu system is unique in that the transverse momentum dependence of the HBT radii
in three dimensions has only been studied at STAR.
This system-wide comparison is clearly important in understanding the physics behind the space-momentum correlations present in each system.

\begin{figure}[]
\resizebox{.4\textwidth}{!}{\includegraphics{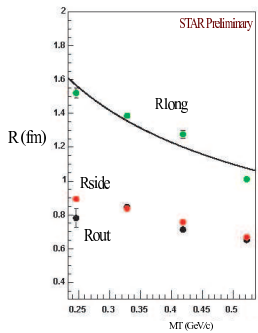}}
\caption{\label{mtdeppp} The transverse mass dependence of the 3D Bertsch-Pratt
radii in pp collisions at STAR. The line guides the eye with a 1/$\sqrt{M_T}$ parameterization.}
\end{figure}

\begin{figure}[!b]
\resizebox{.4\textwidth}{!}{\includegraphics{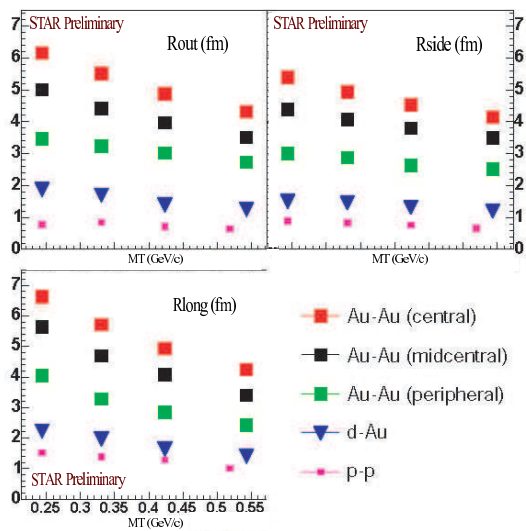}}
\caption{\label{mtdep} The transverse mass dependence of the 3D Bertsch-Pratt
radii in pp, dAu, and AuAu at STAR. Three centralities are displayed for AuAu.  
The pp data points are the same as those shown in Figure~(\ref{mtdeppp}).}
\end{figure}

Figure~(\ref{mtdeppp}) shows the $M_T$ dependence of the three-dimensional Bertsch-Pratt Gaussian
pion radii as seen in pp collisions.  A significant dependence
is seen in $R_{\rm{long}}$, suggestive of a boost-invariant expansion along the beam direction.
In addition, a small, but non-negligible, dependence of the transverse radii is also seen.
In Fig.~(\ref{mtdep}) the same pp data from Fig.~(\ref{mtdeppp}) are plotted 
in conjunction with the AuAu and dAu Gaussian radii (note the change of scale between Fig.~(\ref{mtdeppp}) and (\ref{mtdep})).  
Like in Figure~(\ref{onedall}), the system scale naturally grows smaller as one goes from central AuAu reactions
to dAu and pp collisions.

\begin{figure}[]
\resizebox{.4\textwidth}{!}{\includegraphics{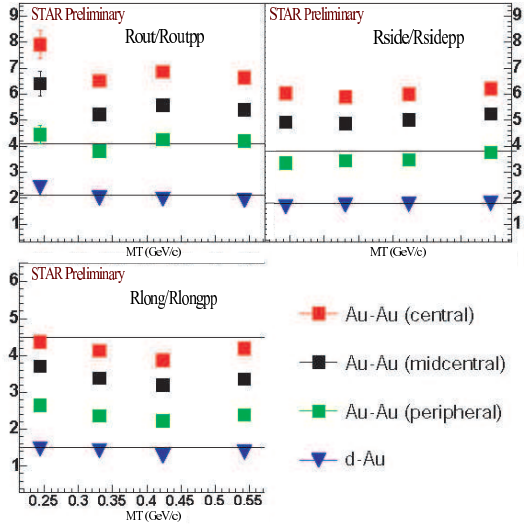}}
\caption{\label{mtdeprat} The ratio of the dAu and AuAu radii to the pp radii as displayed in 
Figure~(\ref{mtdep}).  The lines are provided to guide the eye.}.
\end{figure}

Most interesting is Figure~(\ref{mtdeprat}) which shows the ratio of the AuAu and dAu radii
to the pp radii.  The ratios are strikingly flat, given the very different mechanisms presumably driving these systems.
The similar $R_{\rm{long}}$ dependence is perhaps not unexpected since all of these systems were analyzed at
mid-rapidity, a nearly boost-invariant region at STAR for all systems considered.  
Very different mechanisms, from hydrodynamic expansion to inside-out string
fragmentation, give similar behavior of $R_{\rm{long}}$ versus $M_T$ strictly from boost-invariant arguments.

However,
in the transverse direction the results are more perplexing.  In pp collisions, the transverse dynamics are most likely driven
by the independent fragmentation of (at most) a few strings, some of which will form jets.  In AuAu collisions, the transverse
dynamics are governed by collective expansion such as flow.  
It seems strange that these two mechanisms give rise to very similar
$M_T$ dependence of the transverse radii.  Perhaps much can be learned from this observation and the effect is still under investigation.

There is a long history of using HBT in elementary particle physics to study QCD and the space-time-momentum structure of hadronization.
The multi-system capabilities of RHIC provide a unique laboratory for linking 
what has been historically studied in elementary 
particle collisions and what is known about HBT in heavy ion physics.  By studying the HBT of the pp and dAu systems in the context of AuAu collisions
at STAR, we hope to gain a better understanding of the freeze-out of nuclear matter under various extreme conditions: from hot and dilute
pp collisions, where the space-time-momentum structure of hadronization itself is probed -- to the cooler and denser, highly interacting,
nuclear medium generated in AuAu reactions, where final state collective effects dominate.

%
%
%

\bibliography{QM2004bib}

\begin{thebibliography}{8}
\expandafter\ifx\csname natexlab\endcsname\relax\def\natexlab#1{#1}\fi
\expandafter\ifx\csname bibnamefont\endcsname\relax
  \def\bibnamefont#1{#1}\fi
\expandafter\ifx\csname bibfnamefont\endcsname\relax
  \def\bibfnamefont#1{#1}\fi
\expandafter\ifx\csname citenamefont\endcsname\relax
  \def\citenamefont#1{#1}\fi
\expandafter\ifx\csname url\endcsname\relax
  \def\url#1{\texttt{#1}}\fi
\expandafter\ifx\csname urlprefix\endcsname\relax\def\urlprefix{URL }\fi
\providecommand{\bibinfo}[2]{#2}
\providecommand{\eprint}[2][]{\url{#2}}

\bibitem[{\citenamefont{Goldhaber et~al.}(1960)\citenamefont{Goldhaber,
  Goldhaber, Lee, and Pais}}]{hbtgold60}
\bibinfo{author}{\bibfnamefont{G.}~\bibnamefont{Goldhaber}},
  \bibinfo{author}{\bibfnamefont{S.}~\bibnamefont{Goldhaber}},
  \bibinfo{author}{\bibfnamefont{W.-Y.} \bibnamefont{Lee}}, \bibnamefont{and}
  \bibinfo{author}{\bibfnamefont{A.}~\bibnamefont{Pais}},
  \bibinfo{journal}{Phys. Rev.} \textbf{\bibinfo{volume}{120}},
  \bibinfo{pages}{300} (\bibinfo{year}{1960}).

\bibitem[{\citenamefont{Gyulassy et~al.}(1979)\citenamefont{Gyulassy,
  Kauffmann, and Wilson}}]{hbtgyu79}
\bibinfo{author}{\bibfnamefont{M.}~\bibnamefont{Gyulassy}},
  \bibinfo{author}{\bibfnamefont{S.~K.} \bibnamefont{Kauffmann}},
  \bibnamefont{and} \bibinfo{author}{\bibfnamefont{L.~W.}
  \bibnamefont{Wilson}}, \bibinfo{journal}{Phys. Rev.}
  \textbf{\bibinfo{volume}{C20}}, \bibinfo{pages}{2267} (\bibinfo{year}{1979}).

\bibitem[{\citenamefont{Boal et~al.}(1990)\citenamefont{Boal, Gelbke, and
  Jennings}}]{hbtboal90}
\bibinfo{author}{\bibfnamefont{D.~H.} \bibnamefont{Boal}},
  \bibinfo{author}{\bibfnamefont{C.~K.} \bibnamefont{Gelbke}},
  \bibnamefont{and} \bibinfo{author}{\bibfnamefont{B.~K.}
  \bibnamefont{Jennings}}, \bibinfo{journal}{Rev. Mod. Phys.}
  \textbf{\bibinfo{volume}{62}}, \bibinfo{pages}{553} (\bibinfo{year}{1990}).

\bibitem[{\citenamefont{Heinz and Jacak}(1999)}]{hbthei99}
\bibinfo{author}{\bibfnamefont{U.~W.} \bibnamefont{Heinz}} \bibnamefont{and}
  \bibinfo{author}{\bibfnamefont{B.~V.} \bibnamefont{Jacak}},
  \bibinfo{journal}{Ann. Rev. Nucl. Part. Sci.} \textbf{\bibinfo{volume}{49}},
  \bibinfo{pages}{529} (\bibinfo{year}{1999}).

\bibitem[{\citenamefont{Csorgo and Hegyi}(2000)}]{hbtcso00}
\bibinfo{author}{\bibfnamefont{T.}~\bibnamefont{Csorgo}} \bibnamefont{and}
  \bibinfo{author}{\bibfnamefont{S.}~\bibnamefont{Hegyi}},
  \bibinfo{journal}{Phys. Lett.} \textbf{\bibinfo{volume}{B489}},
  \bibinfo{pages}{15} (\bibinfo{year}{2000}).

\bibitem[{\citenamefont{Brown and Danielewicz}(1997)}]{hbtbro97}
\bibinfo{author}{\bibfnamefont{D.~A.} \bibnamefont{Brown}} \bibnamefont{and}
  \bibinfo{author}{\bibfnamefont{P.}~\bibnamefont{Danielewicz}},
  \bibinfo{journal}{Phys. Lett.} \textbf{\bibinfo{volume}{B398}},
  \bibinfo{pages}{252} (\bibinfo{year}{1997}).

\bibitem[{\citenamefont{Adler}(2001)}]{hbtstar01}
\bibinfo{author}{\bibfnamefont{C.}~\bibnamefont{Adler}, \bibfnamefont{{\em et
  al.}}}, \bibinfo{journal}{Phys. Rev. Lett.} \textbf{\bibinfo{volume}{87}},
  \bibinfo{pages}{082301} (\bibinfo{year}{2001}).

\bibitem[{\citenamefont{Lisa}(2003)}]{hbtlisa03}
\bibinfo{author}{\bibfnamefont{M.~A.} \bibnamefont{Lisa}},
  \bibinfo{journal}{Acta Phys. Polon.} \textbf{\bibinfo{volume}{B35}},
  \bibinfo{pages}{37} (\bibinfo{year}{2003}).

\end{thebibliography}

\end{document}